\documentstyle[12pt]{article}
\newcommand{\be}{\begin{equation}}
\newcommand{\ee}{\end{equation}}
\newcommand{\bea}{\begin{eqnarray}}
\newcommand{\eea}{\end{eqnarray}}
\newcommand{\al}{\alpha}
\newcommand{\bt}{\beta}

\newcommand{\Gm}{\Gamma}
\newcommand{\dl}{\delta}

\newcommand{\kp}{\kappa}
\newcommand{\lm}{\lambda}

\newcommand{\rh}{\rho}

\newcommand{\sg}{\sigma}

\newcommand{\Om}{\Omega}

\begin{document}
\title{On the adiabatic expansion of the visible space 
in a higher dimensional cosmology}
\author{K. Kleidis and D. Papadopoulos\\
{\small Section of Astrophysics, Astronomy and Mechanics}\\
{\small Department of Physics}\\
{\small Aristotle University of Thessaloniki}\\
{\small 54006 Thessaloniki, Greece} }

\maketitle
\begin{abstract}

In the context of higher-dimensional cosmologies we study the conditions under 
which adiabatic expansion of the visible external space is possible, when 
a time-dependent internal space is present. The analysis is based on a 
reinterpretation of the four-dimensional stress-energy tensor in the presence 
of the extra dimensions. This modifies the usual adiabatic energy conservation 
laws for the visible Universe, leading to a new type of cosmological evolution 
which includes large-scale entropy production in four dimensions.

\end{abstract}
\section{Introduction}

The mathematical background for a non-linear gravitational lagrangian theory, 
free from metric derivatives of orders higher than the second, was formulated by 
Lovelock [1], who proposed that the most general gravitational lagrangian is of the 
form
\be
{\cal L} = \sqrt {- g} \sum_{m=0}^{n/2} \lm_m {\cal R}^{(m)}
\ee
where $\lm_m$ are arbitrary constants, $n$ denotes the spacetime dimensions, $g$ is 
the determinant of the metric tensor and ${\cal R}^{(m)}$ are functions of the Riemann 
curvature tensor of the form 
\be
{\cal R}^{(m)} = {1 \over 2^m} \dl_{\al_1...\al_{2m}}^{\bt_1...\bt_{2m}} 
{\cal R}_{\bt_1 \bt_2}^{\al_1 \al_2} ...{\cal R}_{\bt_{2m-1} \bt_{2m}}^{\al_{2m-1} \al_{2m}}
\ee
where $\dl_{\al_1...\al_{2m}}^{\bt_1...\bt_{2m}}$ is the generalized Kronecker 
symbol. In Eq. (2),
${\cal R}^{(1)} = {1 \over 2} {\cal R}$ is the Einstein-Hilbert (EH) lagrangian, while 
${\cal R}^{(2)}$ is a particular combination of the quadratic terms, known as 
the Gauss-Bonnett (GB) combination, since in four dimensions it satisfies the 
functional relation
\be
{\dl \over \dl g^{\mu \nu}} \int \sqrt {-g} \left ( {\cal R}^2 - 4 
{\cal R}_{\mu \nu}{\cal R}^{\mu \nu} + {\cal R}_{\mu \nu \kp \lm}
{\cal R}^{\mu \nu \kp \lm} \right ) d^4 x = 0
\ee
corresponding to the GB theorem [2]. Introduction of this term into the 
gravitational lagrangian will not affect the four dimensional field equations. 
From Eqs. (1) and (3) it becomes evident that if the gravitational lagrangian 
contains terms of the curvature tensor of orders higher than the second $(m \geq 2)$, 
then one needs to have a spacetime of more than four dimensions.

This idea has received much attention 
as a candidate for the unification of all foundamental interactions, including 
gravitation, in the framework of supergravity or in superstrings [3-10]. In most 
higher-dimensional theories of gravity the extra dimensions are assumed 
to form, at the present epoch, a compact manifold ({\em internal space}) of very small 
size compared to that of the three-dimensional visible space ({\em external space}) 
[11] and therefore they are unobservable at the energies currently available. This 
leads to the problem of {\em compactification} of the extra dimensions [12]. It has 
been recently suggested that compactification of the extra space may be achieved, 
in a natural way, by adding a square curvature term, ${\cal R}_{\mu \nu \kp \lm}
{\cal R}^{\mu \nu \kp \lm}$, in the EH action for the gravitational field [13]. In this 
context, the higher-dimensional theories are closely related to those of non-linear 
lagrangians and their combination probably yields to a natural generalization of 
General Relativity (GR).

Many problems of the four-dimensional {\em standard model} attempted 
to be solved in the context of a higher-dimensional gravity theory [14]. Among them,
many effords have been made to explain the creation of the observed entropy content, 
as a result of the dynamic evolution of the internal space [5,7,8,15,16]. The 
four-dimensional Einstein equations are purely 
adiabatic and reversible and, consequently, can hardly provide, by themselves, an 
explanation relating to the origin of cosmological entropy. Therefore, to account for the 
entropy observed in the Universe one has either to assume it as an initial condition 
or to account for it through some dissipative mechanism [17]. This picture radically 
changes in the presence of a time-dependent internal space. Its dynamic evolution, 
preferably its contraction, could release energy in the external space, thus leading 
to an entropy change in four dimensions. This idea, 
although correct in principle, has been able to be applied only in spacetime 
models with a very large number of extra dimensions $(D \sim 40)$ [7] or after 
an unnatural fine-tuning of all the gauge coupling parameters [8].

In the present paper we explore the same idea under a different prespective, 
dealing with the thermodynamics of the visible space in the case where a time -
dependent internal one is present. 
We propose a phenomenological macroscopic approach, using a four-dimensional 
reinterpretation of the higher-dimensional matter stress-energy tensor 
[18-21]. Accordingly, we show that both subspaces do not correspond to 
{\em isolated} thermodynamical 
systems but to {\em closed} ones [22], which allow for energy transfer between them. 
In this respect, a four-dimensional observer comoving with the 
matter content will see an extra amount of "{\em heat}" received by the ordinary 
three-dimensional 
space, which is due entirely to the contraction of the extra dimensions.
Now, adiabatic expansion of the external space occurs only under certain conditions, 
while the irreversible contraction of the internal space could lead to entropy production 
in three dimensions.

\section{Four-dimensional thermodynamics in a higher-dimensional cosmology}

We consider a spacetime of $n = 1 + 3 + D$ dimensions which has been split topologically 
into three homogeneous and isotropic factors, $T \times V^3 \times V^D$, where $T$ is the 
time direction, $V^3$ is the three-dimensional {\em external space}, representing the visible 
Universe and $V^D$ is the D-dimensional {\em internal space} which consists of the extra 
dimensions. The $n$-dimensional line-element is $(\hbar = c = 1)$
\be
ds^2 = - dt^2 + R^2(t) d \sg_{ext}^2 + S^2(t) d \sg_{int}^2
\ee
in which the metrics of the factor spaces are of the form
\be
d \sg_{ext}^2 = {1 \over (1 + k_{ext} {r_{ext}^2 \over 4} )^2} \sum_{i=1}^3 (d x_{ext}^i)^2
\ee
and
\be
d \sg_{int}^2 = {1 \over (1 + k_{int} {r_{int}^2 \over 4} )^2} \sum_{i=1}^D (d x_{int}^i)^2
\ee
with
$$
r_{ext}^2 = \sum_{i=1}^3 (x_{ext}^i)^2 \; \; \; and \; \; \; r_{int}^2 = \sum_{i=1}^D (x_{int}^i)^2.
$$
In what follows we consider only models of an already compactified internal space, 
i.e. we examine the process of its contraction. The contraction of the inner 
dimensions presupposes their separation from the ordinary ones and strictly speaking starts 
immediately after compactification [10]. As far as the factor metric (6) is concerned, 
compactification may be achieved either in terms of a $D$-dimensional sphere, with 
$k_{int} = 1$ or in terms of a $D$-dimensional torus, with $k_{int} = 0$.

Euler variation of Eq. (1) gives the Lovelock tensor ${\cal L}_{\mu \nu}$ [1] 
(Greek indices refer to the whole $n$-dimensional spacetime). ${\cal L}_{\mu \nu}$ 
is the most 
general symmetric and divergenceless tensor which describes the propagation of the 
gravitational field and is a function of the metric tensor and its first and 
second order derivatives. In this case, the generalized gravitational field 
equations read
\be
{\cal L}_{\mu \nu} = - 8 \pi G_n \; T_{\mu \nu}
\ee
where 
\be
{\cal L}_{\nu}^{\mu} = {1 \over 2^m} \sum_{m=0}^{n/2} \dl_{\nu \al_1...\al_{2m}}^
{\mu \bt_1...\bt_{2m}} 
{\cal R}_{\bt_1 \bt_2}^{\al_1 \al_2} ...{\cal R}_{\bt_{2m-1} \bt_{2m}}^
{\al_{2m-1} \al_{2m}}
\ee
is the Lovelock tensor, $G_n$ is the $n$-dimensional gravitational constant and 
$T_{\mu \nu}$ is the energy-momentum tensor, which is also included in the field 
equations through an action principle and contains all the matter and the energy 
present in the spacetime region $V_n$. The maximally symmetric character of the 
two subspaces in the cosmological model (4) restricts the form of $T_{\mu \nu}$ 
which, in this case, is diagonal and may be considered as representing a 
$n$-dimensional fluid. There exists one common energy density,
\be
T^{00} = \rh
\ee
corresponding to the one time direction and two different pressures, $p_{ext}$ and 
$p_{int}$, associated with each factor space respectively. The spatial pressures 
do not necessarily coincide with the corresponding thermodynamical quantities. 
Both $p_{ext}$ and $p_{int}$ are isotropic in each factor space separately. 
In addition, since both subspaces are homogeneous and isotropic the 
energy density and the associated pressures are functions of the cosmic time only.

We consider a $n$-dimensional observer who is locally at rest, i.e. comoving 
with the fluid along the world lines with a tangent velocity vector of the form
$$
u^{\mu} = \left ( 1, 0, ... , 0 \right )
$$
Then, any $3+D$-dimensional proper comoving volume of fluid, $V_{3+D} = R^3S^D$, 
may be considered as an isolated thermodynamical system which, accordingly, 
evolves adiabatically [20,22]. Hence, there exist two equations 
of state in the form $p_d = p_d(\rh)$, one for each subspace of dimensions $d$ [23]. 
Once the overall matter-energy density $\rh$ is properly defined, the form 
of the associated pressures, $p_{ext}$ and $p_{int}$, may be directly obtained in 
terms of those equations [10]. In four dimensions the general linear equation 
of state 
\be
p = \left ( {m \over 3} - 1 \right ) \; \rh 
\ee
covers most of the matter components considered to fill the Early Universe, like 
quantum vacuum $(m = 0)$, gas of strings $(m=2)$, dust $(m=3)$, radiation $(m=4)$ 
and Zel'dovich ultrastiff matter $(m=6)$. A mixture of such components obeys the 
expansion law 
\be
\rh \; = \; \sum_m {M_m \over R^m}
\ee
where $M_m$ is constant if no transitions between the different components occur 
(there are no dissipative mechanisms) [10]. The generalization of Eq. (10) to 
multidimensional models which consist of factor spaces, requires [10]
\be
p_d = \left ( {m_d \over d} - 1 \right ) \rh 
\ee
where $d$ is the number of dimensions of the corresponding factor space. For the 
cosmological model under consideration, $d=3$ for the external space and $d=D$ 
for the internal one. In this case the evolution of $\rh$ results in
\be
\rh \; = \; \sum_{m_3} \: \sum_{m_D} \; {M_{m_3, m_D} \over R^{m_3} S^{m_D}}
\ee
where again, in the absence of dissipation, $M_{m_3, m_D}$ is constant.

A very interesting property of ${\cal L}_{\mu \nu}$ is that it is divergenceless, 
${\cal L}^{\mu \nu} \;_{; \mu} = 0$. This condition imposes, through the field 
equations (7), that the same is also true for the energy-momentum tensor of the 
fluid source
\be
T^{\mu \nu} \; _{; \mu} = 0
\ee
The specific form of the metric tensor together with the fact that $\rh, p_{ext}$ 
and $p_{int}$ are functions only of the cosmic time, restricts the number of 
components of Eq.(14) leaving only one which is not satisfied identically. 
Namely $T^{0 \mu} \; _{; \mu} = 0$. It leads to the condition
\be
{\dot \rh} + 3 \left ( \rh + p_{ext} \right ) { {\dot R} \over R} + 
D \left ( \rh + p_{int} \right ) { {\dot S} \over S} = 0
\ee
where a dot denotes derivative with respect to the cosmic time, t.

Eq.(15) describes the interchange of energy between matter and gravitation, in a 
curved spacetime [24] and corresponds to the energy conservation law for an observer 
who is comoving with the fluid. In this respect, $\rh, p_{ext}$ and $p_{int}$ are 
the matter-energy density and pressures {\em locally measured} by a $n$-dimensional 
observer inside the proper comoving volume $V_{3+D} = R^3S^D$. Eq.(15) may be 
written in a more convenient form, as follows
\be
{d \over dt} \rh = {1 \over R^3} \left [ {d \over dt}(\rh R^3) \; + 
\; p_{ext} {d \over dt}(R^3) \right ] 
+ {1 \over S^D} \left [ {d \over dt}(\rh S^D) \; + \; p_{int} 
{d \over dt}(S^D) \right ]
\ee

The question that arises now, is what will see a four-dimensional observer who is 
unaware of the existence of the extra dimensions probably due to the fact that 
the "physical size" of the internal space is very small. As regards the external 
space, the four dimensional observer may also choose a coordinate system in which he is 
locally at rest, comoving with a four-dimensional projection of the fluid element. Both 
comoving coordinate systems are of the same origin. 

In order to answer this question we need to determine how the $3+D$-dimensionally 
defined quantities $\rh$ and $p_{ext}$ are "projected" onto the three-dimensional 
spatial section of the external space. In the context of higher-dimensional 
cosmologies, when a {\em compact} internal space is present, the three-dimensional 
matter-energy density of the external space is defined [8,25] as the integral of 
the overall $3+D$-dimensional matter-energy density $\rh$ over the proper volume 
of the $D$-dimensional (closed and bounded) internal space. That is
\be
\rh_3 = \int_{V_D} \rh \; \sqrt {g_D} \; d^D x
\ee
This statement is in complete agreement with the definition of the {\em 
stress-energy tensor on a hypersurface} considered in Ref. 26 (Eq. 21.163, p.552). It 
expresses the ability of a four-dimensional observer to measure, at each time, 
the total energy of the (multidimensional) Universe [25] which is justified by 
the fact that the extra dimensions usually manifest themselves in the 
four-dimensional energy-momentum tensor [27-30]. In this case, all the energy 
included in the extra space is somehow being {\em "projected"} onto the 
three-dimensional hypersurface which forms the external space. The total amount 
of matter-energy in a general $3+D$-dimensional proper comoving volume is given by
\be
{\cal E} = \int_{V_{3+D}} \rh \; \sqrt {-{\hat g}} \; d^{3+D} x
\ee
where ${\hat g}$ is the determinant of the overall metric tensor. Since the 
metric (4) is diagonal, Eq.(18) is decomposed to
\be
{\cal E} = \int_{V_{3+D}} \rh \; \sqrt {-g} \; \sqrt {g_D} \; d^{3+D} x 
= \int_{V_3} d^3 x \sqrt {-g} \left ( \int_{V_D}  \rh \; \sqrt {g_D} \; d^D x 
\right )
\ee
where $g$ is the determinant of the four-dimensional external space and $g_D$ 
is the corresponding determinant of the $D$-dimensional internal one.
According to what previously stated, we define an effective three-dimensional 
total energy of the form 
\be
{\cal E}_3^{eff} = \int_{V_3} \rh_3 \; \sqrt {-g} \; d^3 x
\ee
introducing ${\cal E}_3^{eff} = {\cal E}$. Then, combination of Eqs. (19) and 
(20) implies that Eq.(17) holds. However, $\rh$ depends only on the cosmic time 
and therefore $ \rh_3 = \rh V_D$. For the cosmological model under consideration, 
as regards any proper comoving volume of the internal space, $V_D \sim S^D$, Eq.(17) 
may be written in the form
\be
\rh_3 = \rh \: S^D
\ee
where in Eq.(21) we have ignored factors which appear in the formula for $V_D$ 
of the form $(2 \pi)^{(D+1)/2}/ \Gm ( {D+1 \over 2} )$ a thing that amounts to 
a redefinition of $S(t)$.

In this case, $\rh_3$ may be identified as the {\em "phenomenological matter-energy 
density"}, measured inside a three-dimensional proper comoving volume of fluid in the visible 
space $(V_3 \sim R^3)$, when a compact time-dependent internal space is present. We see that the 
presence of the extra dimensions modifies the physical content of the ordinary 
Universe (a not unexpected result e.g. see also [8,18,19,25,27-30]). In the absence 
of the extra dimensions ($D = 0$) we obtain $\rh_3 = \rh$ corresponding to the energy 
density of a perfect fluid source in four dimensions [25].

When a corresponding analysis is carried out for the energy density of the internal 
space, from Eq.(19) we obtain
\be
\rh_D = \rh \: V_3 = \rh \: R^3
\ee

To account for the "phenomenological" expression of the three-dimensional pressure, 
$p_3$, in the presence of the inner dimensions, we see from Eq.(12) that the 
pressure associated to the external space is
$$
p_{ext} = \left ( {m_3 \over 3}  - 1 \right ) \; \rh 
$$
which by virtue of Eq.(21) reads
\be
p_{ext} = \left ( {m_3 \over 3} - 1 \right ) \; {1 \over S^D} \: \rh_3 \; 
\Leftrightarrow p_{ext} S^D = \left ( {m_3 \over 3} - 1 \right ) \rh_3 
\ee
According to Eq.(10), a four-dimensional observer unaware of the existence of the 
extra dimensions would recognize the r.h.s. of Eq.(23) as representing the "physical 
pressure" in four dimensions. Therefore, combination of Eqs. (10) and (24) implies 
\be
p_3 = p_{ext} \: S^D
\ee
The quantity $p_3$ corresponds to the phenomenological 
"physical" pressure, measured by an observer inside a proper comoving volume of the 
external space, in the presence of a time-dependent internal space. However, as 
we will see later on, the expansion of the external space in this case is no longer 
adiabatic. Therefore, variation of entropy occur in four dimensions and, together 
with Eq.(10) one should impose an additional equation of state [23], of the form 
$ {\cal E}_3 = {\cal E}_3 ({\cal S}_3) $, 
where ${\cal S}_3$ is the total entropy in the external space. In the absence of the 
extra dimensions ${\cal S}_3 = constant$, $p_3 = p_{ext}$ and therefore $T_{\mu \nu}$ 
would represent a four-dimensional perfect fluid. 

As regards the corresponding "physical" pressure of the internal space, we obtain
\be
p_D = p_{int} \: R^3
\ee
By virtue of Eqs. (21), (24) and (25), Eq. (16) is written in the form 
\be
d \left (\rh_3 R^3 \right ) \; + \; p_3 \: d \left ( R^3 \right ) \; = \; 
- \: p_D \: d \left ( S^D \right )
\ee
Eq.(26), which represents the $n$-dimensional conservation law (15), is in 
complete correspondance with the first law of thermodynamics for a {\em closed 
system} in four-dimensions (any arbitrary comoving volume of fluid), dealing 
with changes in the energy content between succesive states of 
equilibrium [20,21,26,31,32]. This consideration leads to an extension of thermodynamics 
as associated with the four-dimensional cosmology. 

To address this statement in detail let us, at first, consider the case of a 
static internal space, $S = S_0$, as it is imposed to be the case at the present 
epoch. Now, Eq.(26) becomes
\be
d \left (\rh_3 R^3 \right ) \; + \; p_3 \: d \left ( R^3 \right ) \; = \; 0  
\ee
Eq.(27) is used to describe the adiabatic evolution of a proper comoving volume 
element of the visible Universe [31], when the curvature is small enough, so 
that the self-energy density of the gravitational field is much lower than the 
matter-energy density [32]. Equivalently, it corresponds to the first thermodynamical 
law for the adiabatic evolution of an {\em isolated} system [20-22,26,28,31,32]. 
Then, we may interpret $\rh_3$ and $p_3$ as the 
{\em true} thermodynamical energy-density and pressure [8,20,25]. The case in 
which the internal space is static corresponds to an {\em adiabaticity condition} 
in four-dimensions. 

The situation is not the same when a time-dependent internal space is 
present. Then the first thermodynamical law in four dimensions is given by Eq.(26) 
and consequently the expansion of a proper comoving volume of fluid 
in the external space is no longer adiabatic. 

The same is also true for any proper comoving volume in the internal space. Indeed, 
inserting Eqs. (22), (24) and (25) into Eq.(16) we obtain
\be
d \left (\rh_D S^D \right ) \; + \; p_D \: d \left ( S^D \right ) \; =\; - \: 
p_3 \: d \left (R^3 \right )
\ee
Eq.(28) represents the first law of thermodynamics as regards a proper comoving 
volume of fluid in $D$ dimensions. It is worth to note that in the case of a static 
internal space Eq.(28) is also reduced to Eq.(27). 

We see that, although Eq.(16) states that the evolution of the $4+D$-dimensional 
spacetime is isentropic and the total mass-energy is conserved, when it is 
reduced to the 
four-dimensional or $D$-dimensional expression, indicates that the two factor 
spaces do not constitute isolated systems but {\em closed} ones, which permits for energy 
transfer between them [22]. In this case an extra amount of "heat" 
is received by the four-dimensional system, which is due entirely to the evolution 
of the internal space. Namely
\be
d \left (Q_{ext} \right ) \; = \; -\: p_D \: d \left ( S^D \right )
\ee
The same is also true for the corresponding system in $D$ dimensions
\be
d \left (Q_{int} \right ) \; = \; - \: p_3 \: d \left ( R^3 \right )
\ee
Since the $4+D$-dimensional spacetime evolves adiabaticaly we impose 
\be
d \left ( Q_{ext} \right ) \; = \; - \: d \left ( Q_{int} \right )
\ee
Eq.(31) reduces to the condition
\be
{1 \over R^{m_3} S^{m_D}} \: R^3 S^D = constant
\ee
which, according to Eq.(13), states that the total energy in a $4+D$-dimensional 
comoving volume remains constant.

In what follows we are interested in studying the thermodynamics of the external 
space since it represents the ordinary Universe. Therefore, we mainly use Eqs. (26) 
and (29). A subsequent study may also be performed for any proper comoving volume 
of fluid in the internal space, if we use Eqs. (28) and (30) in the place of Eqs. 
(26) and (29). 

\section{Entropy production due to cosmological contraction of the extra 
dimensions}

The extra amount of energy received by the external 
space will increase the random microscopic motions within a comoving volume element 
of fluid in the ordinary Universe. In this respect it modifies the 
four-dimensional energy density $\rh_3$ which now, according to Eqs. (13) and (21) 
in the case of one-component matter, reads
\be
\rh_3 \; \sim \; {1 \over R^{m_3} \: S^{m_D-D}}
\ee
and a similar formula holds also for the corresponding pressure, $p_3$. 
Therefore, it also modifies the evolution of the visible Universe through 
the corresponding cosmological field equations [20-22].

However, when $p_{int} = 0$ (i.e. when $m_D = D$) the r.h.s. of Eq.(29) is zero. 
In this case, 
condition (27) for adiabatic expansion of the external space is satisfied, 
no matter which the dynamic behaviour of the internal space might be. Hence, 
the case of a pressureless internal space also corresponds to an adiabaticity 
condition in four dimensions. Since $d Q_{ext} = 0$ the existence of the 
internal space does not imply any energy contributions to the external one. 
No energy contributions means that there is not any change in the evolution of 
the external space which could be caused due to the existence of the extra dimensions 
[in connection see Eq.(33)]. This result was recently observed in the level of the 
field equations of a five-dimensional quadratic cosmology, which, for $p_{int} = 0$ 
decouple [33]. In this case, $\rh_3 \sim R^{-\: m_3}$ and the external space 
expands under its own laws of 
evolution. Therefore the two subspaces are completely {\em disjoint}. 

When $p_{int} \neq 0$ the two factor spaces are not disjoint at all. If the 
internal space is filled with a conventional type of matter for which $m_D \geq D$, 
like dust $(m_D = D)$, radiation $(m_D = {4 \over 3} D) $ or ultrastiff matter 
$(m_D = 2D)$, from Eq.(12) we obtain $p_{int} > 0$. In this case, Eq.(29) indicates 
that when the internal space contracts $dQ_{ext} > 0$, i.e. energy is transfered 
to the external space from the extra dimensions. In the same fashion, expansion 
of the internal space corresponds to $dQ_{ext} < 0$ and therefore, enegry is 
extracted from the external space and transfered to the internal one in order to 
maintain its expansion. The opposite results are obtained if the internal space 
is filled with unconventional types of matter, for which $m_D < D$, like quantum 
vacuum $(m_D = 0)$ or gas of strings $(m_D = {2 \over 3} D )$, since in this case 
$p_{int} < 0$. In what follows we always consider that $m_D \geq D$.

By virtue of Eq.(12), the first thermodynamical law (26) may be written in the form 
\be
d \left ( {\cal E}_3(t) \right ) \; + \; p_3 \: d \left ( R^3 \right ) \; = \; 
{\cal E}_3(t) \: d \left ( ln S^{D-m_D} \right )
\ee
Eq.(34) indicates that the external space may be energy-supported by the evolution 
of the extra dimensions which acts as a source of {\em internal energy} in four 
dimensions. This energy transfer stops when the internal space becomes static 
$(dQ_{ext} = 0)$. For $p_{int} \neq 0$, the visible Universe evolves 
adiabatically only after stabilization of the inner dimensions. In any other case, 
the extra amount of "heat" on the r.h.s. of Eq.(34) 
corresponds to an entropy change, $d{\cal S}_3 \neq 0$, in four dimensions
\be
T_3 (t) \: d \left ( {\cal S}_3 (t) \right ) \; = \; {\cal E}_3 (t) \: d \left ( 
ln S^{D-m_D} \right )
\ee
The second law of thermodynamics states that $d{\cal S}_3 > 0$. Therefore, according to 
Eq.(35), for $m_D > D$ there is only one possible evolution of the internal space after 
compactification, that is, its contraction. In the context of higher-dimensional theories 
where only conventional matter is present, the contraction of the internal space 
appears to be an irreversible process, since the reverse one is thermodynamically 
forbidden.

In principle we can integrate Eq.(35) to obtain the total amount of entropy produced 
inside a causal volume in four dimensions. The limits of integration range from an 
initial time $t_{in}$ at which compactification (i.e. separation of the extra 
dimensions from the ordinary ones) begins, up to the final stage $t_f$ at which 
stabilization of the internal space is achieved. However, there still remains the 
question of how to calculate the unknown function $T_3(t)$, since for ${\cal S}_3 \neq 0$ 
we can not use the definition of {\em thermodynamic temperature} (e.g. see 
[31,34]) for a fluid source in four dimensions. It is more convenient to find 
the form of ${\cal S}_3(t)$ as a function of $S(t)$. We do so by noting that, 
when the entropy in the external space varies, Eq.(10) is not by itself sufficient 
to determine completely the state of the matter content [22, 23]. We also need 
a second equation of state in the form ${\cal E}_3 = {\cal E}_3 ({\cal S}_3)$. 
When a {\em closed} thermodynamical system evolves non-adiabatically, 
the {\em free energy of Helmholtz} $F = {\cal E} - T {\cal S}$ equals to the {\em 
generalized thermodynamical potential} $\Om = -pV$ [35]. For the corresponding system 
inside a proper comoving volume of the external space we obtain [31]
\be
{\cal S}_3(t) \; = \; {R^3 \over T_3(t)} \: \left ( \rh_3 \: + \: p_3 \right )
\ee
since $\rh_3$ and $p_3$ are functions of the cosmic time only. Eq.(36) gives 
the entropy associated to the measured thermodynamical content of the external space 
at each time. For adiabatic expansion of the external space the r.h.s. of Eq.(36) 
is constant. Combining Eqs. (23) and (36), for $m_3 \neq 0$, we obtain
\be
{\cal E}_3(t) \; = \; {3 \over m_3} \: T_3(t) \: {\cal S}_3(t)
\ee
Eq.(37) corresponds to the second equation of state which, together with 
$p = p(\rh)$, is appropriate for the description of a thermodynamical 
system during non-adiabatic procedures [23,31,34,35]. Inserting Eq.(37) 
into Eq.(35) we finally obtain
\be
{\cal S}_3(t) \; = \; {\cal S}_{30} \: \left ( {S \over S_0} \right )^{{3 \over 
m_3}\:(D - m_D)}
\ee
where ${\cal S}_{30}$ is the constant value of entropy of the external space from 
the moment at which $S(t)$ becomes static and afterwards, while $S_0$ represents 
the value of the internal scale factor at stabilization. Eq.(38) determines the 
entropy produced in the external space due to the dynamical evolution 
of the extra dimensions. ${\cal S}_3$ depends on three free 
parameters of the theory, namely {\bf (i)} the equation of state in the external 
space through $m_3$, {\bf (ii)} the equation of state in the internal space 
through $m_D$ and {\bf (iii)} the number of the extra dimensions. For $m_D > D$ 
the entropy of the external space increases as the internal space contracts.

Once the form of ${\cal S}_3(t)$ is found as a function of the internal scale 
factor $S(t)$, the corresponding expression of $T_3(t)$ inside a proper comoving 
volume of fluid in four dimensions results from the combination of Eqs. (37) and 
(38). For $m_3 \neq 0$, it can be cast into the form 
\be
T_3(t) \; = \; T_{30} \: {1 \over R^{m_3-3}} \: \left ( {S \over S_0} 
\right )^{- \: {1 \over m_3} \: (m_D-D) \: (m_3-3)}
\ee
where $T_{30}$ is constant. Eq.(39) indicates that for any of the adiabaticity 
conditions: {\bf (i)} the internal space is static, $S = S_0$ or 
{\bf (ii)} $m_D = D$ (i.e. $p_{int} = 0$), the functional form of the 
external temperature is reduced to the corresponding expression of the 
four-dimensional FRW cosmology
\be
T_3(t) \sim { 1 \over R^{m_3-3}}
\ee
For non-adiabatic expansion of the external space, when both 
subspaces are filled with conventional types of matter, i.e. $m_3 \geq 3$ 
and $m_	D \geq D$, contraction of the internal space results in an increase 
of the four-dimensional temperature and therefore, "heat" is received by 
the external space. This result verifies the corresponding considerations 
of Abbott et al. [7] and Kolb et al. [8].

\section{Discussion and Conclusions}

In the present paper we have examined the conditions under which adiabatic 
evolution of the visible space is possible in the context of a higher-dimensional 
non-linear theory of gravity. 
We have considered a $n$-dimensional cosmological model $(n = 1+3+D)$, 
consisting of one time direction and two homogeneous and isotropic factor 
spaces, the external space and the internal one. The Universe is filled with 
one-component matter, in the form of fluid. 
In this model there is one common energy density, $\rh$ and two different 
pressures, $p_{ext} $ and $p_{int}$, associated to each factor space respectively.

To impose the conditions on the adiabatic evolution of the external space we have 
used a reinterpretation of the $n$-dimensional stress-energy tensor from the point 
of view of a four-dimensional observer who is unaware of the existence of the extra 
dimensions. For a compact internal space, the three-dimensional energy-density, 
$\rh_3$, has been defined considering that all the energy included in the extra 
space is projected, at each time, onto the spatial section of the visible 
Universe [8,25,26]. Then, $\rh_3$ is given by Eq.(21), while similar expressions hold 
for the pressures, Eqs. (24) and (25).
Using these expressions we have written the $n$-dimensional first thermodynamical law 
(15) in the four-dimensional point of view (26). In this case, the two subspaces 
correspond to {\em closed thermodynamical systems} which allows for energy transfer 
between them. As regards the external space, in contrast to four-dimensional 
cosmology, $dQ_{ext} \neq 0$ and therefore its evolution is no longer adiabatic. 

According to Eq.(29), for $p_{int} > 0$, contraction of the inner dimensions implies 
$dQ_{ext} > 0$, i.e. extra energy is received by the external space, while expansion 
of the inner dimensions implies $dQ_{ext} < 0$, i.e. energy is extracted from the 
external space. Since the $n$-dimensional spacetime corresponds to an {\em isolated 
thermodynamical system} this energy amount is subsequently received by the internal 
space in order to maintain its expansion. There are only two cases at which the 
evolution of the visible space is adiabatic: {\bf (i)} The case of a static internal 
space, $S(t) = S_0$ and {\bf (ii)} the case of a pressureless internal space, 
$p_{int} = 0$. Both cases correspond to {\em adiabaticity conditions} since 
$dQ_{ext} = 0$. Now, the two subspaces are completely disjoint and the cosmological 
field equations decouple [33].

For non-adiabatic evolution of the visible Universe, the extra amount of energy 
received by the external space results to a large-scale entropy production 
$(d{\cal S}_3 \neq 0)$ in four dimensions. Taking into account the second 
thermodynamical law $d{\cal S}_3 > 0$, we see that for $p_{int} > 0$ is true only 
when the internal space contracts. In this context, contraction of the extra 
dimensions appears to be an irreversible process, since the reverse one is 
thermodynamically forbidden. Both the entropy produced in the visible space 
due to the contraction of the extra dimensions ${\cal S}_3 (t)$ and the 
corresponding external 
temperature $T_3(t)$, have been obtained as functions of the internal scale 
factor, Eqs. (38) and (39). These expressions increase as the {\em physical 
size} of the internal space decreases and they depend on three free parameters 
of the theory: {\bf (i)} The equation of state in the external space, through 
$m_3$, {\bf (ii)} the equation of state in the internal space, through $m_D$ 
and {\bf (iii)} the number of the extra dimensions, $D$. It is probable that a 
fine-tunning of these parameters could lead to production 
of a considerably large entropy amount, inside a causal volume in four dimensions, 
to match up with the observational data [7,8,31].

In order to maintain the treatment as general as possible we have not imposed 
anything, throughout this article, about the form of the {\em "external heat"} 
$dQ_{ext} \neq 0$, i.e. the amount of energy which is 
received by the visible Universe due to the dynamic evolution of the internal space. 
Recent developments indicate that the cosmological contraction of 
one extra dimension could lead to massless particle production (radiation) in 
the ordinary space [36]. In 
this case, the energy received by the external space probably corresponds to that of 
the produced radiation which, in turn, is extracted from the anisotropic 
gravitational field [37-40]. Moreover, in the context of the {\em open 
thermodynamical systems} [22] it has been also shown that irreversible particle 
production in four dimensions is closely related to entropy creation, 
$d{\cal S}_3 > 0$, in the ordinary space [20,40]. An extension of these results to 
higher-dimensional cosmologies, together with a detailed study of the possible 
relation between them, would be very interesting and it will be the scope 
of a future work.

\vspace{1.cm}

{\bf Acknowledgements:}  The authors would like to express their gratitude to 
Professor N. K. Spyrou for his comments and his advices during many helpful 
discussions. One of us (K. K.) would like to thank the Greek State Scholarships 
Foundation for the financial support during this work.
\\

\section*{References}

\begin{itemize}

\item[1]Lovelock D., J. Math. Phys. 12, 498 (1971)
\item[2]Kobayashi S. and Nomizu K. {\em Foundations of Differential Geometry II}, 
Wiley Interscience, N. Y. (1969).
\item[3]Chodos A. and Detweiler S., Phys. Rev. D 21, 2167 (1980)
\item[4]Freund P. G. O., Nucl. Phys. B 209, 146 (1982)
\item[5]Alvarez E. and Belen-Gavela M., Phys. Rev. Lett. 51, 931 (1983)
\item[6]Randjbar-Daemi S., Salam A. and Strathdee J., Phys. Lett. 135B, 388 (1984)
\item[7]Abbott R., Barr S. and Ellis S. D., Phys. Rev. D 30, 720 (1984)
\item[8]Kolb E. W., Lindley D. and Seckel D., Phys. Rev. D 30, 1205 (1984)
\item[9]Demaret J. and Hanquin J. L., Phys. Rev. D 31, 258 (1985)
\item[10]Bleyer U., Liebscher D. E. and Polnarev A. G., Class. Quantum Grav. 8, 477 (1991)
\item[11]Green M. B., Schwartz J. H. and Witten E. {\em Superstring Theory}, Vols 1, 2, 
Cambridge Univ. Press, Cambridge (1987)
\item[12]Lee H. C. {\em Introduction to Kaluza - Klein Theories}, Ed. by H. C. Lee, 
World Scientific, Singapore (1984)
\item[13]Muller-Hoissen F. and Stuckl H., Class. Quantum Grav.5, 27 (1988)
\item[14]Applequist T., Chodos A. and Freund P. {\em Modern Kaluza - Klein Theories}, 
Addison - Wesley (1987)
\item[15]Barr S. and Brown L., Phys. Rev D 29, 2779 (1984)
\item[16]Sahdev D., Phys. Lett. 137B, 155 (1984)
\item[17]Misner S. W. ApJ 151, 431 (1968)
\item[18]Davidson A. and Owen D. A., Phys. Lett. 155B, 247 (1985)
\item[19]Ibanez J. and Verdaguer E., Phys. Rev. D 34, 1202 (1986)
\item[20]Prigogine I., Geheniau J., Gunzig E. and Nardone P., Gen. Relativ. Gravit. 
21, 767 (1989)
\item[21]Sudharsan R. and Johri V. B., Gen. Relativ. Gravit. 26, 41 (1994)
\item[22]Prigogine I. {\em Thermodynamics of Irreversible Processes}, Wiley, New 
York (1961)
\item[23]Ryan M. Jr. and Shepley L. C. {\em Homogeneous Relativistic Cosmologies}, 
Princeton University Press, Princeton New Jersey (1975) 
\item[24]Weinberg S. {\em Gravitation and Cosmology} (p. 158), Wiley, New York (1972)
\item[25]Farina-Busto L., Phys. Rev. D 38, 1741 (1988)
\item[26]Misner C. W., Thorne K. S. and Wheeler J. A. {\em Gravitation}, Freeman, 
San Francisco (1973)
\item[27]Wesson P. S. Phys. Lett. B 276, 299 (1992)
\item[28]Wesson P. S. Ap J 394, 19 (1992)
\item[29]Chatterjee S. and Sil A., Gen. Relat. Gravit. 25, 307 (1993)
\item[30]Mashhoon B., Liu H. and Wesson P. S., Phys. Lett. B 331, 305 (1994)
\item[31]Narlikar J. V. {\em Introduction to Cosmology}, Jones and Bartlett 
Publishers Inc., Boston (1983)
\item[32]Landau L. D. and Lifshitz E. M. {\em The Classical Theory of Fields}, 
Pergamon Press, London (1974)
\item[33]Kleidis K. and Papadopoulos D., (preprint) submitted to J. Math. Phys.,
JMP 5-824 (1995) 
\item[34]Geroch R., J. Math. Phys. 36, 4226 (1995)
\item[35]Mandl F. {\em Statistical Physics}, Wiley and Sons Ltd., Chichester (1974)
\item[36]Garriga J. and Verdaguer E., Phys. Rev D 39, 1072 (1989)
\item[37]Brout R., Englert F. and Gunzig E., Ann. Phys. 115, 78 (1978)
\item[38]Brout R., Englert F. and Spindel P., Phys. Rev. Lett. 43, 417 (1979)
\item[39]Birrell N. D. and Davies P. C. W. {\em Quantum Fields in Curved Space}, 
Cambridge University Press, Cambridge (1982)
\item[40]Nesteruk A. Class. Quantum Grav. 8, L241 (1991)

\end{itemize}

\end{document}